%% file: acl_latex.tex
\pdfoutput=1
\documentclass[11pt]{article}

\usepackage[final]{acl}

\usepackage{times}
\usepackage{latexsym}

\usepackage[T1]{fontenc}
\usepackage[utf8]{inputenc}

\usepackage{microtype}

\usepackage{inconsolata}

\usepackage{graphicx}

\usepackage{pgfplots}
\pgfplotsset{compat=1.16}
\usepackage{pgfplotstable}
\usepackage{filecontents}
\usepackage{multirow}
\usepackage{geometry}
\usepackage{booktabs}
\usepackage{arydshln}

\input{math_commands}

\title{Streamlining Conformal Information Retrieval via Score Refinement}

\author{Yotam Intrator\thanks{Equal Contribution.} \\
  \texttt{yotami@google.com}\And
  Regev Cohen$^\ast$ \\
  \texttt{regevcohen@google.com}  \And
  Ori Kelner$^\ast$ \\
  \texttt{orikelner@google.com} \AND
  Roman Goldenberg \\
  \texttt{gnamor@gmail.com}  \And
  Ehud Rivlin \\
  \texttt{ehudr@cs.technion.ac.il}  \And
  Daniel Freedman \\
  \texttt{danielfreedman@gmail.com} \AND
  Verily AI (Google Life Sciences), Israel.
  }

\begin{document}
\maketitle
\begin{abstract}
\input{sections/abstract}

\end{abstract}

% \thanks{NON-ARCHIVAL submission}

\section{Introduction}
\label{sec:intro}
\input{sections/intro}

\section{Background}
\label{sec:background}
\input{sections/background}

\section{Method}
\label{sec:method}
\input{sections/method}

\section{Experiments}
\label{sec:experiments}
\input{sections/experiments}

\section{Conclusion}
\label{sec:conclusion}
\input{sections/conclusion}

\section{Limitations}
\label{sec:limit}
\input{sections/limitations}

% \section{Broader Impact}
% \label{sec:impact}
% \input{sections/impact}

% \section*{Acknowledgments}

% Bibliography entries for the entire Anthology, followed by custom entries
% \bibliography{anthology,custom}
% Custom bibliography entries only
\bibliography{custom}

\clearpage
\appendix
\section{Additional Experiments}
\label{sec:appendix}
\input{sections/appendix}

\end{document}

%% file: math_commands.tex
\usepackage{amsmath,amsfonts,bm}
\usepackage{amssymb}
\usepackage{amsthm}
\usepackage[inline]{enumitem}
\usepackage{placeins}

\def\eqref#1{equation~\ref{#1}}
\def\1{\bm{1}}

\DeclareMathAlphabet{\mathsfit}{\encodingdefault}{\sfdefault}{m}{sl}
\SetMathAlphabet{\mathsfit}{bold}{\encodingdefault}{\sfdefault}{bx}{n}

\def\gC{{\mathcal{C}}}
\def\gD{{\mathcal{D}}}

\def\gP{{\mathcal{P}}}
\def\gQ{{\mathcal{Q}}}

\def\gS{{\mathcal{S}}}

\def\sP{{\mathbb{P}}}

\newcommand{\E}{\mathbb{E}}

\newcommand{\parbasic}[1]{\noindent \textbf{#1} \hspace{0.5mm}}

%% file: sections/abstract.tex
% Information retrieval methods are fundamental to modern applications, such as Retrieval Augmented Generation (RAG), which enhances the accuracy of large language models by incorporating external knowledge sources.
% While promising, IR methods typically lack statistical guarantees, prompting the development of conformal prediction for retrieving sets guaranteed to include relevant information with user-specified probabilities.
% Existing conformal approaches often produce excessively large retrieved sets, leading to increased computational costs and slow response times.
% In this work, we introduce a rescoring method that applies a simple monotone transformation, inspired by ranking measures, to the retrieval system's scores. By applying standard conformal prediction methods to these transformed scores, we achieve significantly smaller retrieved sets while maintaining their statistical guarantees. Our experiments on [Fill in] validate the effectiveness of our approach, demonstrating its efficiency in producing compact sets containing the relevant information when compared to competing approaches.https://braintex.goog/project/66b45ea624c22900a89823bb

% Short Verstion
Information retrieval (IR) methods, like retrieval augmented generation, are fundamental to modern applications but often lack statistical guarantees. Conformal prediction addresses this by retrieving sets guaranteed to include relevant information, yet existing approaches produce large-sized sets, incurring high computational costs and slow response times. In this work, we introduce a score refinement method that applies a simple monotone transformation to retrieval scores, leading to significantly smaller conformal sets while maintaining their statistical guarantees. Experiments on various BEIR benchmarks validate the effectiveness of our approach in producing compact sets containing relevant information.

%% file: sections/intro.tex
Information retrieval (IR) methods lie at the heart of numerous modern applications, ranging from search engines and recommendation systems to question-answering platforms and decision support tools. These methods facilitate the identification and extraction of relevant information from vast collections of data, enabling users to access the knowledge they seek efficiently and effectively. 
A popular example of IR is Retrieval Augmented Generation (RAG), a technique for reducing hallucinations in large language models (LLMs) by grounding their responses on factual information retrieved from external sources. 

While IR methods have been widely adopted, they traditionally lack statistical guarantees on the relevance of retrieved information. This limitation can lead to uncertainty regarding the reliability and correctness of the retrieved information. Conformal prediction~\citep{angelopoulos2021gentle, angelopoulos2021learn} is an uncertainty quantification framework that can be used with any underlying model to
construct sets that are statistically guaranteed to contain the ground truth with a user-specified probability. Conformal prediction has expanded far beyond its initial classification focus~\citep{vovk2005algorithmic, angelopoulos2021gentle, ringel2024early}, now encompassing diverse applications like regression, image-to-image translation~\citep{angelopoulos2022image, kutiel2023conformal}, and foundation models~\citep{gui2024conformal}, advancing to enable control of any monotone risk function~\citep{angelopoulos2022conformal}.
In the context of IR, recent methods~\citep{xu2024conformal,li2023trac, angelopoulos2023recommendation} have incorporated conformal prediction into ranked retrieval systems to ensure the reliability and quality of retrieved items. 
However, existing conformal methods often produce excessively large retrieved sets, implying high computational costs and slower response times.

\begin{figure*}[t]
    \centering
    \includegraphics[width=0.9\linewidth]{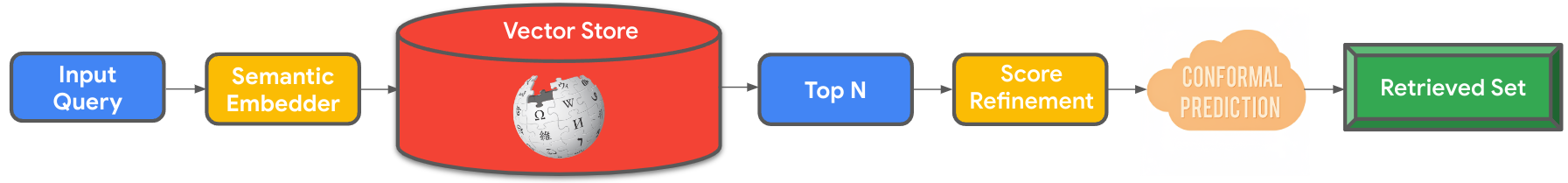}
    
    \caption{Retrieval Pipeline. The query is first embedded using a semantic embedder, and then the top $N$ candidates are retrieved from a vector store. Crucially, their corresponding scores then undergo a refinement transformation before being passed through a conformal prediction method that outputs an adaptive set of documents.}
    \label{fig:conformal_pipeline}
\end{figure*}

In this work, we address this limitation by introducing a novel score refinement method that employs a simple yet effective monotone transformation, inspired by ranking measures, to adjust the scores of any given information retrieval system.
By applying standard conformal prediction methods to these refined scores, we deliver significantly smaller retrieved sets while preserving their statistical guarantees, striking a crucial balance between efficiency and accuracy. An illustration of the proposed pipeline is shown in Figure~\ref{fig:conformal_pipeline}.
We validate the effectiveness of our method through experiments on three of BEIR~\citep{thakur2021beir} benchmark datasets, demonstrating its ability to outperform competing approaches in producing compact sets that contain the relevant information.

%% file: sections/background.tex
To lay the groundwork for our work, we present a simplified description of the operation of information retrieval systems and how conformal inference can be seamlessly integrated within this context.

\subsection{Information Retrieval: Overview}
Consider a large information database $\mathcal{D} = \{d_1, d_2, ..., d_N\}$. At inference time, an IR model $R:\gQ\rightarrow \gD$ accepts a query $q\in\gQ$ as input and returns a subset of candidates $\gS\subset \gD$. To do this, the IR model computes a semantic embedding $e_q = E(q)$ for the query and compares it to pre-computed embeddings $e_i=E(d_i)$ for each item in the database using a similarity metric:
\begin{equation}
    s_i = \texttt{sim}(e_q, e_i),
\end{equation}
where $E$ is the chosen representation model (e.g., a neural network encoder) and $\texttt{sim}$ is a similarity metric, such as cosine similarity. Subsequently, the items are typically ranked based on their similarity scores, and the top ranked items are retrieved, forming the following set
\begin{equation}
    \gS_K \triangleq \Big\{d_i \in \gD : s_i \geq s_{(K)}\Big\}
    \label{eq:topk}
\end{equation}
where $s_{(K)}$ denotes the $K$th largest similarity score, for a predefined $K> 0$ constant across all queries. 

The approach above suffers from two key limitations. First, using a fixed $K$ can be problematic: it might be too restrictive for some queries, leading to the omission of relevant information, while for others, it might be too permissive, resulting in the retrieval of numerous redundant or irrelevant items. The latter scenario significantly impacts efficiency and prolongs response times. Second, this approach lacks guarantees that truly relevant information, such as a specific item $d^*$ within the database $\gD$, will be included in the retrieved set $S$.

\subsection{Conformal Prediction for Retrieval}

Conformal prediction can be seamlessly integrated into IR systems by constructing calibrated prediction sets designed to include, on average, the desired information with a user-specified high probability. Formally, given a query $q$ and its corresponding similarity scores $s_i$, we construct a prediction set parameterized by $\tau > 0$ as follows:
\begin{equation}
    \gC_\tau(q) \triangleq \{d_i \in \gD : c_i \leq \tau \},
\end{equation}
where $c_i \triangleq -s_i$ represents a \textit{non-conformity} score.
To appropriately set the value of $\tau$, we utilize a held-out calibration dataset $\gD_C$ consisting of $n$ samples $(q_i, d_i) \in \gQ \times \gD$ drawn exchangeably from an underlying distribution $\gP$. Here, $q_i$ represents a query whose most relevant information is assumed to be a single item $d_i$ from the database, for simplicity. 
Given a user-chosen error rate $\alpha \in [0, 1]$, we set $\tau$ as the $\frac{(n+1)(1-\alpha)}{n}$-th quantile of the calibration non-conformity scores. This ensures that for a new exchangeable test sample $(q_{n+1}, d_{n+1})$, we have the following marginal coverage guarantee:
\begin{equation}
    \sP\big(d_{n+1} \in \gC_\tau(q_{n+1})\big) \geq 1 - \alpha
\end{equation}
for any distribution $P$. The probability above is marginal (averaged) over all $n + 1$ calibration and test samples. This ensures that the IR model retrieves sets of adaptive size, guaranteed to contain the relevant information at least $\alpha$-fraction of the time, thereby overcoming the limitation above.

While the conformal sets above use a calibrated threshold, other parameterizations are possible, such as setting the calibration parameter to the set size $K$, as in (\ref{eq:topk}). Furthermore, it is important to note that the description above merely presents conformal prediction in its simplest, most common form. However, there have been significant advancements in the field in recent years, leading to the development of more involved and efficient conformal methods \citep{romano2020classification, angelopoulos2020uncertainty} and to extensions that provide guarantees beyond marginal coverage~\citep{angelopoulos2022conformal, fisch2020efficient, li2023trac}.

% vovk~\citep{vovk2005algorithmic}
% \newline
% aps romano~\citep{romano2020classification}
% \newline
% raps~\citep{angelopoulos2020uncertainty}
% \newline
% conformal prediction + retrieval - (Efficient Conformal Prediction via Cascaded Inference with Expanded Admission) - vovk on the first stage - \citep{fisch2020efficient}
% \newline
% conformal rag - vovk on on the retrieval~\citep{li2023trac}

%% file: sections/method.tex
Integrating conformal prediction to IR systems enhances their reliability by providing statistical guarantees. However, CP methods prioritize trustworthiness and are not optimized for efficiency, thus they often produce excessively large retrieval sets.

Following the above, our goal is to improve the predictive efficiency of CP methods by reducing the average size of the retrieved sets $\E_q\big[|\gC_\tau(q)|\big]$, while maintaining their coverage guarantees.
In contrast to approaches that focus on improving the IR model or developing more efficient conformal methods, we propose an alternative approach that introduces an intermediate step of score refinement. Specifically, given a query $q$ and its scores $\gS = \{s_1, s_2, \dots, s_N\}$, we adjust them prior to employing conformal prediction $T(\gS) = \{t_1, t_2, \dots, t_N\}$.

In designing the transformation $T$, we identify that scores from different queries can vary significantly in scale. This can cause the calibration threshold $\tau$ to be dominated by queries with small scores, leading to excessively large prediction sets.
To mitigate this issue, we first normalize the retrieval scores by dividing them by their maximum, ensuring that scores across all queries are comparable in scale. We remark that the maximum score $s_{\max}$ can be interpreted as the IR model's confidence. When this value is small, it suggests a lack of relevant information for the given query, suggesting that ideally no items should be retrieved. Thus, normalization in such scenarios may be counterproductive, resulting in irrelevant items. However, we assume the corpus is sufficiently extensive to contain at least one relevant item for any query, an assumption particularly valid for the calibration.

Next, assume without loss of generality that the scores are sorted in decreasing order: $\gS = \{s_{(1)}, s_{(2)}, \dots, s_{(N)}\}$, where $s_{(r)}$ is the $r$th largest score and $r\geq 1$ represents its rank.
Inspired by ranking measures~\citep{yining2013theoretical}, we define our transformation as follows
\begin{equation}
    T(s_{(r)}, r)\triangleq\frac{s_{(r)}}{s_{\max}}W(r)
\end{equation}
where $W(r)\in[0,1]$ is a discount function that penalizes scores based on their rank. We specifically employ the inverse logarithm decay $W(r)=\frac{1}{\log(1+r)}$, which offers a balance between emphasis on top items and exploration of lower-ranked items. To offer additional flexibility, we introduce a hyperparameter  $\lambda\in[0, 1]$:
\begin{equation}
    T(s_{(r)}, r)\triangleq\frac{s_{(r)}}{s_{\max}}\frac{1}{\log(1+r^{\lambda})}.
\end{equation}
We tune $\lambda$ by performing a search over a sequence of values to minimize the set size on a validation set.
Note the transformation is monotone, preserving the IR model's induced order and maintaining its core functionality. Furthermore, it is simple to implement, computationally efficient, and easily integrated into existing systems.
As demonstrated in the following section, the proposed transformation is highly effective in reducing the size of the conformal retrieved sets.

% \subsection{Problem Formulation}

% \subsection{Rescoring}

%% file: sections/experiments.tex
\subsection{Setup}
\label{subsec:setup}

% beir benchmark~\citep{thakur2021beir}
% \newline
% scifact~\citep{wadden2020fact}
% \newline
% fever~\citep{thorne2018fever}
% \newline
% fiqa~\citep{maia201818}
% \newline
% bge~\citep{bge_embedding}
% \newline
% e5 mistral~\citep{wang2023improving}
% \newline
% mistral 7b~\citep{jiang2023mistral}
% \newline

\parbasic{Datasets}
For our evaluation, we utilized BEIR~\citep{thakur2021beir}, a large collection of information retrieval benchmarks. Specifically, we focus on the following datasets: FEVER~\citep{thorne2018fever}, SCIFACT~\citep{wadden2020fact}, and FIQA~\citep{maia201818}. Data statistics are presented at Table~\ref{table:data_stats}. 
% Due to the absence of a designated development set in SCIFACT, the test set was randomly partitioned, resulting in 150 queries allocated to both the newly created calibration set and the test set. 
It is important to note that each query within these datasets may have multiple relevant documents within the corpus. For this study, we adopted a pragmatic approach, considering the document with the highest score among the relevant documents as the ground truth. This ensures that a successful retrieval implies at least one relevant document is present in the inference set.
% assuming that any single correct matching document is sufficient to contain the answer. Therefore, during calibration and testing, if at least one ground truth document is present in the inference set, it is considered a successful retrieval. 

To simulate real-world production environments, we employ a vector store, specifically FAISS-GPU~\citep{johnson2019billion} for its efficiency and performance in handling large-scale databases. We  retrieve the top $2,000$ documents for each query and apply our refinement process exclusively to these initially retrieved documents.  

\input{tables_figures/data_description}

\parbasic{Embedders}
Initial semantic scores were derived using deep sentence embedders, which encode textual input into a fixed-dimensional latent space where semantic similarity is represented by vector proximity.
We employ two models: BGE-large-1.5~\citep{bge_embedding} (326M parameters) and E5-Mistral-7b model~\citep{wang2023improving} (7B parameters). BGE-large-1.5 is a smaller model with a latent representation dimension of 1024, whereas E5-Mistral, a finetuned encoder version of the mistral-7b model, has a latent representation dimension of 4096.
The semantic score between a query $q$ and a candidate document $d$ is the cosine similarity between their respective latent representations. 

% The cosine similarity is defined as:
% \begin{equation}
%     \text{Cosine Similarity}(q, d) \triangleq \frac{q \cdot d}{\norm{q} \cdot\norm{d}}
% \end{equation}
% where $q \cdot d$ denotes the dot product of the latent representations, and $\norm{q}$ and $\norm{d}$ are the magnitudes of the latent representations.

\parbasic{Competitors}
For our method, we employ the Vanilla CP method~\citep{vovk2005algorithmic}, applying it to the refined retrieval scores. We compared our approach to three established approaches: \textit{Baseline}, which applies Vanilla CP directly to the retrieval scores without modification; \textit{TopK}, which utilizes Vanilla CP but calibrates to a fixed set size $K$ for all queries; \textit{APS}~\citep{romano2020classification} and \textit{RAPS}~\citep{angelopoulos2020uncertainty}, which introduce novel conformity scores.

\input{tables_figures/lambda_figure}
\input{tables_figures/fever_figure}

\subsection{Results}
\label{subsec:results}
We first conduct experiments on the smaller SCIFACT dataset to optimize the hyperparameter $\lambda$. The results, shown in Figure~\ref{fig:lambda_ablation}, reveal a favorable value for $\lambda$, prompting us to set $\alpha=0.05$.  

Next, we conduct experiments on the large-scale FEVER dataset. As illustrated in Figure~\ref{fig:conformal_comparison}, our score refinement method consistently outperforms other approaches by producing significantly smaller retrieved sets in experiments with BGE-large-1.5 across various values of $\alpha$, while maintaining comparable, albeit slightly lower, empirical coverage.
Results for the other datasets are summarized in Table~\ref{tab:results_fiqa_scifact}, consistent with our previous findings. We note that RAPS produced comparable results to APS, so we omit them for brevity. Additional results using E5-Mistral, which exhibit similar trends, are presented in Table~\ref{tab:results_mistral} of the appendix, along with an ablation study comparing other simple transformations.

%% file: tables_figures/data_description.tex
\begin{table}[h]
    \centering
    \small

    \begin{tabular}{cccc}
        \toprule
        \textbf{Dataset} & \textbf{\#Corpus} & \textbf{\#Calibration} & \textbf{\#Test} \\
        \midrule
        FEVER & 5,416,568 & 6,666 & 6,666 \\
        SCIFACT & 5,183 & 150$^*$ & 150$^*$ \\
        FIQA & 57,638 & 500 & 648 \\
        \bottomrule
        
    \end{tabular}
    \caption{Data Summary. \#Corpus indicates the number of documents, while \#Calibration and \#Test indicate the number of queries. As SCIFACT lacks a calibration set, we randomly split its test set into calibration and test subsets.}
\label{table:data_stats}
\end{table}

%% file: tables_figures/lambda_figure.tex
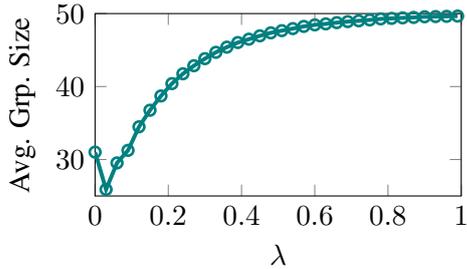
\begin{figure}
    \centering
    \begin{tikzpicture}
        \begin{axis}[
            width=0.4\textwidth,
            height=0.25\textwidth,
            xlabel={$\lambda$},
            ylabel={Avg. Grp. Size},
            xmin=0, xmax=1,
            ymin=25, ymax=50,
            xtick={0,0.2,...,1},
            % ytick={1,1.1, ..., 2},
            % grid=both,
            % major grid style={line width=.2pt,draw=gray!50},
            % minor grid style={line width=.1pt,draw=gray!10},
            % legend style={at={(1,1.05)}, anchor=south east}
        ]
        \addplot[
            color=teal, % Changed color
            mark=o, % Changed marker
            mark options={fill=white, line width=1pt}, % Style for markers
            line width=1.5pt % Thicker line
        ]
        coordinates {
            % (0.0, 1.89)
            % (0.03, 1.71)
            % (0.06, 1.67)
            % (0.09, 1.68)
            % (0.12, 1.71)
            % (0.15, 1.72)
            % (0.18, 1.72)
            % (0.21, 1.74)
            % (0.24, 1.76)
            % (0.27, 1.79)
            % (0.3, 1.82)
            % (0.33, 1.82)
            % (0.36, 1.86)
            % (0.39, 1.88)
            % (0.42, 1.90)
            % (0.45, 1.92)
            % (0.48, 1.92)
            % (0.51, 1.93)
            % (0.54, 1.95)
            % (0.5700000000000001, 1.95)
            % (0.6, 1.97)
            % (0.63, 1.97)
            % (0.66, 1.98)
            % (0.6900000000000001, 1.99)
            % (0.72, 1.99)
            % (0.75, 1.99)
            % (0.78, 2.00)
            % (0.81, 2.00)
            % (0.84, 2.00)
            % (0.87, 2.00)
            % (0.9, 2.00)
            % (0.93, 2.00)
            % (0.96, 2.00)
            % (0.99, 2.00)
            
            (0.0, 31.01)
            (0.03, 25.89)
            (0.06, 29.53)
            (0.09, 31.25)
            (0.12, 34.48)
            (0.15, 36.75)
            (0.18, 38.71)
            (0.21, 40.40)
            (0.24, 41.75)
            (0.27, 42.85)
            (0.3, 43.82)
            (0.33, 44.69)
            (0.36, 45.40)
            (0.39, 46.03)
            (0.42, 46.49)
            (0.45, 46.95)
            (0.48, 47.36)
            (0.51, 47.66)
            (0.54, 47.94)
            (0.5700000000000001, 48.24)
            (0.6, 48.45)
            (0.63, 48.59)
            (0.66, 48.76)
            (0.6900000000000001, 48.90)
            (0.72, 49.01)
            (0.75, 49.13)
            (0.78, 49.27)
            (0.81, 49.34)
            (0.84, 49.41)
            (0.87, 49.47)
            (0.9, 49.56)
            (0.93, 49.59)
            (0.96, 49.62)
            (0.99, 49.66)

        };
        % \addlegendentry{Data}
        \end{axis}
    \end{tikzpicture}
    \vspace{-10pt}
    \caption{Impact of $\lambda$ value on average group size using 
    BGE-large-1.5 on SCIFACT with $\alpha$ = 0.05.}
    \label{fig:lambda_ablation}
\end{figure}

%% file: tables_figures/fever_figure.tex
\definecolor{naive_color}{RGB}{66,133,244}
\definecolor{aps_color}{RGB}{234,67,53}
\definecolor{topk_color}{RGB}{251,188,4}
% \definecolor{naive_normalized_color}{RGB}{148,55,180}
\definecolor{ours_color}{RGB}{52,168,82}

\begin{figure}
    \centering
    % First plot: Empirical Coverage
    \begin{tikzpicture}
        \begin{axis}[
        legend style={font=\small, nodes={scale=0.7, transform shape}},
            width=0.4\textwidth,
            height=0.27\textwidth,
            ybar=0pt,
            bar width=6pt,
            symbolic x coords={0.9, 0.95, 0.97},
            xtick=data,
            ymin=0.85, ymax=1,
            ylabel={Emp. Cov.},
            xlabel={Desired Coverage (1-$\alpha$)},
            enlarge x limits=0.2,
            % nodes near coords,
            every node near coord/.append style={font=\small},
            legend pos=north west,
            legend style={font=\small},
            % grid=major,
            % ymajorgrids=true,
            % xmajorgrids=true,
            % Avoid multiple squares in legend
            legend image code/.code={
                \draw[#1,fill=#1] (0cm,-0.1cm) rectangle (0.25cm,0.15cm);
            },
        ]
            \addplot+[ybar, fill=naive_color] coordinates {(0.9,0.904) (0.95,0.945) (0.97,0.967)}; \addlegendentry{Baseline}
            \addplot+[ybar, fill=aps_color] coordinates {(0.9,0.87) (0.95,0.929) (0.97,0.953)}; \addlegendentry{APS}
            \addplot+[ybar, fill=topk_color] coordinates {(0.9,0.926) (0.95,0.951) (0.97,0.965)}; \addlegendentry{TopK}
            \addplot+[ybar, fill=ours_color] coordinates {(0.9,0.87) (0.95,0.924) (0.97,0.952)}; \addlegendentry{Ours}
        \end{axis}
    \end{tikzpicture}
    \hfill
    % Second plot: Average Size
    \begin{tikzpicture}
        \begin{axis}[
            width=0.4\textwidth,
            height=0.27\textwidth,
            ybar=0pt,
            bar width=6pt,
            symbolic x coords={0.9, 0.95, 0.97},
            xtick=data,
            ymin=1, ymax=10,
            ylabel={Avg. Grp. Size},
            xlabel={Desired Coverage (1-$\alpha$)},
            enlarge x limits=0.2,
            % nodes near coords,
            every node near coord/.append style={font=\small},
            legend pos=north west,
            legend style={font=\small},
            % grid=major,
            % ymajorgrids=true,
            % xmajorgrids=true,
            % Avoid multiple squares in legend
            legend image code/.code={
                \draw[#1,fill=#1] (0cm,-0.1cm) rectangle (0.25cm,0.15cm);
            },
        ]
            \addplot+[ybar, fill=naive_color] coordinates {(0.9,4.81) (0.95,9.47) (0.97,15.63)};
            \addplot+[ybar, fill=aps_color] coordinates {(0.9,1.52) (0.95,2.13) (0.97,3.15)};
            \addplot+[ybar, fill=topk_color] coordinates {(0.9,2) (0.95,3) (0.97,4)};
            \addplot+[ybar, fill=ours_color] coordinates {(0.9,1.18) (0.95,1.67) (0.97,2.37)};
        \end{axis}
    \end{tikzpicture}
    \caption{Performance comparison using BGE-large-1.5 on FEVER dataset across various values of $\alpha$.}
    \label{fig:conformal_comparison}
\end{figure}
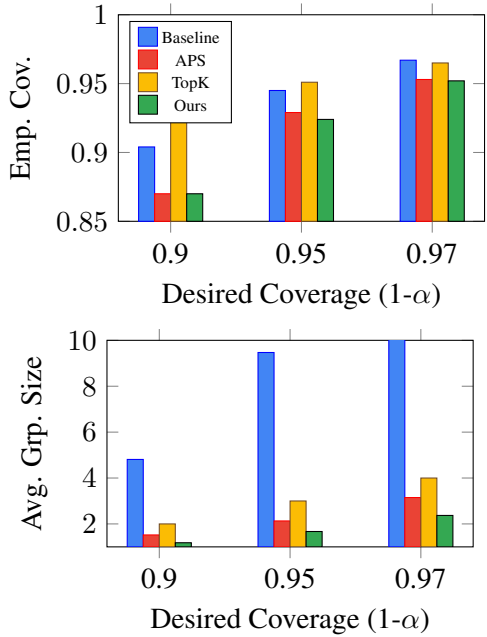

%% file: sections/conclusion.tex
% In this work, we addressed the challenge of large prediction sets in conformal prediction for information retrieval. We introduced a novel score refinement method that applies a simple transformation to the retrieval scores, thereby improving the efficiency of subsequent conformal predictors. 
% Our experiments on the BEIR benchmark, specifically with the FEVER, SCIFACT, and FIQA datasets, demonstrated the effectiveness of our approach in generating compact prediction sets while maintaining strong statistical guarantees. By striking a balance between efficiency and accuracy, our method enables the deployment of conformal prediction in real-world IR systems, enhancing their reliability without sacrificing performance. 

We addressed the challenge of large prediction sets in conformal prediction for IR by introducing a novel score refinement method. Our experiments on the BEIR benchmark demonstrated its effectiveness in generating compact, statistically reliable prediction sets, enabling the deployment of conformal prediction in real-world IR systems without sacrificing performance.

\input{tables_figures/scifact_fiqa_results}

%% file: tables_figures/scifact_fiqa_results.tex
\begin{table}
\centering
\scriptsize
\begin{tabular}{cp{0.2cm}cp{1.5cm}p{2.04cm}}
\toprule
\textbf{Dataset} & \textbf{$\alpha$} & \textbf{Method} & \textbf{Emp. Cov.} & \textbf{Avg. Grp. Size} \\

\midrule
\multirow{4}{*}{FIQA} & \multirow{4}{*}{0.1} & Baseline & 0.89 & 417.77 \\
& & APS & 0.89 & 119.76 \\
& & TopK & 0.87 & 90.0 \\
& & Ours & 0.86 & \textbf{56.72} \\
\hdashline
\multirow{4}{*}{} & \multirow{4}{*}{0.05} & Baseline & 0.94 & 846.0 \\
& & APS & 0.94 & 477.27 \\
& & TopK & 0.92 & 259.0 \\
& & Ours & 0.92 & \textbf{190.5} \\
\hdashline
\multirow{4}{*}{} & \multirow{4}{*}{0.03} & Baseline & 0.96 & 1206.93 \\
& & APS & 0.98 & 1393.96 \\
& & TopK & 0.94 & 480.0 \\
& & Ours & 0.95 & \textbf{347.16} \\
\midrule
\multirow{4}{*}{SCIFACT} & \multirow{4}{*}{0.1} & Baseline & 0.91 & 231.17 \\
& & APS & 0.91 & 30.82 \\
& & TopK & 0.91 & 31.0 \\
& & Ours & 0.85 & \textbf{14.07} \\
\hdashline
\multirow{4}{*}{} & \multirow{4}{*}{0.05} & Baseline & 0.97 & 760.75 \\
& & APS & 0.92 & 91.23 \\
& & TopK & 0.92 & 91.0 \\
& & Ours & 0.89 & \textbf{29.59} \\
\hdashline
\multirow{4}{*}{} & \multirow{4}{*}{0.03} & Baseline & 0.98 & 1211.11 \\
& & APS & 0.95 & 276.15 \\
& & TopK & 0.95 & 279.0 \\
& & Ours & 0.97 & \textbf{160.66} \\
\bottomrule
\end{tabular}
\caption{Performance comparison using BGE-large-1.5 on FIQA and SCIFACT across various values of $\alpha$.}
\label{tab:results_fiqa_scifact}
\end{table}

% \begin{table}
% \centering
% \small
% \begin{tabular}{c|l|c|c}
% \toprule
% \textbf{Dataset} & \textbf{Method} & \textbf{Emp. Cov.} & \textbf{Avg. Grp. Size} \\
% \midrule
% \multirow{4}{*}{FIQA} & Baseline & 0.89/0.94/0.96 & 417.77/846.0/1206.93 \\
% & APS & 0.89/0.94/0.98 & 119.76/477.27/1393.96 \\
% & TopK & 0.87/0.92/0.94 & 90.0/259.0/480.0 \\
% & Ours & 0.86/0.92/0.95 & \textbf{56.72}/\textbf{190.5}/\textbf{347.16} \\
% \hdashline
% \multirow{4}{*}{SCIFACT} & Baseline & 0.91/0.97/0.98 & 231.17/760.75/1211.11 \\
% & APS & 0.91/0.92/0.95 & 30.82/91.23/276.15 \\
% & TopK & 0.91/0.92/0.95 & 31/91/279.0 \\
% & Ours & 0.85/0.89/0.97 & \textbf{14.07}/\textbf{29.59}/\textbf{160.66} \\
% \bottomrule
% \end{tabular}
% \caption{Performance comparison using BGE-large-1.5 on FIQA and SCIFACT across various values of $\alpha$ (0.1, 0.05, 0.03).}
% \label{tab:results_fiqa_scifact}
% \end{table}

%% file: sections/limitations.tex
The conclusions of this study could be further strengthened by evaluating the method on a wider range of datasets and employing diverse embedding models.  Currently, our method does not handle cases where no relevant information exists in the database, potentially limiting its applicability.  Additionally, while we introduced a simple transformation, exploring more involved or even parameterized functions, e.g. neural networks, could further enhance efficiency and statistical guarantees.

% some theortical propeties of the refined scores. 

%% file: sections/appendix.tex
Here evaluate our method with the E5-Mistral embedder on SCIFACT and FIQA datasets. Results, presented in Table~\ref{tab:results_mistral}, show our method consistently outperforms competitors. Moreover, using E5-Mistral leads to improved performance in both empirical coverage and average group size compared to BGE-large-1.5.

In addition to the aforementioned experiments, we compared our method against alternative transformations: \textit{Max Score}, where scores are normalized by dividing each by the maximum score, and \textit{Z-Score}, which standardizes the initial retrieved scores. The results, summarized in Table \ref{tab:app_exp}, show that our score refinement transformation outperforms these other refinement methods in the vast majority of cases.

\input{tables_figures/mistral_scifact_fiqa_results}

\input{tables_figures/ablation_vovk_bge}

%% file: tables_figures/mistral_scifact_fiqa_results.tex
% \begin{table*}
% \centering
% \resizebox{1.02\textwidth}{!}{ % Adjusts table to fit within the text width
% \begin{tabular}{c c c c c c c c c c c c c c c c c}
% \toprule
% \multirow{3}{*}{\textbf{Dataset}} & \multirow{3}{*}{\textbf{$\alpha$}} & \multicolumn{2}{c}{\textbf{Baseline}} & \multicolumn{2}{c}{\textbf{APS}} & \multicolumn{2}{c}{\textbf{TopK}} & \multicolumn{2}{c}{\textbf{Ours}} \\
% \cmidrule(lr){3-4} \cmidrule(lr){5-6} \cmidrule(lr){7-8} \cmidrule(lr){9-10}
%  &  & \textbf{Emp. Cov.} & \textbf{Avg. Grp. Size} & \textbf{Emp. Cov.} & \textbf{Avg. Grp. Size} & \textbf{Emp. Cov.} & \textbf{Avg. Grp. Size} & \textbf{Emp. Cov.} & \textbf{Avg. Grp. Size} \\
% \midrule

% \multirow{3}{*}{SCIFACT} & 0.10 & 0.91 & 68.91 & 0.94 & 17.46 & 0.95 & 19.0 & 0.93 & \textbf{15.09} \\
%  & 0.05 & 0.96 & 311.73 & 0.98 & 139.36 & 0.99 & 150.0 & 0.97 & \textbf{48.71} \\
%  & 0.03 & 0.99 & 1093.85 & 1.0 & 324.09 & 1.0 & 368.0 & 1.0 & \textbf{127.29} \\
% \midrule

% \multirow{3}{*}{FIQA} & 0.10 & 0.91 & 144.31 & 0.9 & 46.48 & 0.89 & 38.0 & 0.9 & \textbf{33.35} \\
%  & 0.05 & 0.96 & 458.79 & 0.95 & 123.09 & 0.94 & 108.0 & 0.94 & \textbf{67.21} \\
%  & 0.03 & 0.98 & 710.86 & 0.97 & 439.64 & 0.96 & 193.0 & 0.96 & \textbf{143.76} \\

% \bottomrule
% \end{tabular}
% }
% \caption{Empirical coverage and average group size for FIQA and SCIFACT, alpha values, and methods using the e5-mistral-7b-instruct.}
% \label{tab:results_mistral}
% \end{table*}

\begin{table}[b!]
\centering
\resizebox{\columnwidth}{!}{ % Adjusts table to fit within the column width
\begin{tabular}{c c c c c}
\toprule
\textbf{Dataset} & \textbf{$\alpha$} & \textbf{Method} & \textbf{Emp. Cov.} & \textbf{Avg. Grp. Size} \\
\midrule

\multirow{12}{*}{SCIFACT} & \multirow{4}{*}{0.10} & Baseline & 0.91 & 68.91 \\
 &  & APS & 0.94 & 17.46 \\
 &  & TopK & 0.95 & 19.0 \\
 &  & \textbf{Ours} & 0.93 & \textbf{15.09} \\
\cmidrule(lr){2-5}

 & \multirow{4}{*}{0.05} & Baseline & 0.96 & 311.73 \\
 &  & APS & 0.98 & 139.36 \\
 &  & TopK & 0.99 & 150.0 \\
 &  & \textbf{Ours} & 0.97 & \textbf{48.71} \\
\cmidrule(lr){2-5}

 & \multirow{4}{*}{0.03} & Baseline & 0.99 & 1093.85 \\
 &  & APS & 1.0 & 324.09 \\
 &  & TopK & 1.0 & 368.0 \\
 &  & \textbf{Ours} & 1.0 & \textbf{127.29} \\
\midrule

\multirow{12}{*}{FIQA} & \multirow{4}{*}{0.10} & Baseline & 0.91 & 144.31 \\
 &  & APS & 0.9 & 46.48 \\
 &  & TopK & 0.89 & 38.0 \\
 &  & \textbf{Ours} & 0.9 & \textbf{33.35} \\
\cmidrule(lr){2-5}

 & \multirow{4}{*}{0.05} & Baseline & 0.96 & 458.79 \\
 &  & APS & 0.95 & 123.09 \\
 &  & TopK & 0.94 & 108.0 \\
 &  & \textbf{Ours} & 0.94 & \textbf{67.21} \\
\cmidrule(lr){2-5}

 & \multirow{4}{*}{0.03} & Baseline & 0.98 & 710.86 \\
 &  & APS & 0.97 & 439.64 \\
 &  & TopK & 0.96 & 193.0 \\
 &  & \textbf{Ours} & 0.96 & \textbf{143.76} \\

\bottomrule
\end{tabular}
}
\caption{Empirical coverage and average group size for FIQA and SCIFACT, alpha values, and methods using the e5-mistral-7b-instruct.}
\label{tab:results_mistral}
\end{table}

%% file: tables_figures/ablation_vovk_bge.tex
\begin{table}[t!]
\centering
\small
\resizebox{\columnwidth}{!}{
\begin{tabular}{c c c c c c}
\toprule
\textbf{Dataset} & \textbf{$\alpha$} & \textbf{Method} & \textbf{Emp. Cov.} & \textbf{Avg. Grp. Size} \\
\midrule

\multirow{12}{*}{FEVER} & \multirow{4}{*}{0.10} & Baseline & 0.90 & 4.81 \\
 & & Max Score & 0.87 & 1.19 \\
 & & Z-Score & 0.85 & 1.63 \\
 & & \textbf{Ours} & 0.87 & \textbf{1.18} \\
 \cmidrule(lr){2-5}
 & \multirow{4}{*}{0.05} & Baseline & 0.95 & 9.47 \\
 & & Max Score & 0.93 & 1.89 \\
 & & Z-Score & 0.92 & 2.44 \\
 & & \textbf{Ours} & 0.93 & \textbf{1.67} \\
 \cmidrule(lr){2-5}
 & \multirow{4}{*}{0.03} & Baseline & 0.97 & 15.63 \\
 & & Max Score & 0.96 & 2.88 \\
 & & Z-Score & 0.95 & 3.28 \\
 & & \textbf{Ours} & 0.95 & \textbf{2.37} \\
\midrule

\multirow{12}{*}{SCIFACT} & \multirow{4}{*}{0.10} & Baseline & 0.91 & 231.17 \\
 & & Max Score & 0.83 & 20.68 \\
 & & Z-Score & 0.88 & 22.01 \\
 & & \textbf{Ours} & 0.85 & \textbf{14.07} \\
 \cmidrule(lr){2-5}
 & \multirow{4}{*}{0.05} & Baseline & 0.97 & 760.75 \\
 & & Max Score & 0.86 & 31.01 \\
 & & Z-Score & 0.91 & 52.91 \\
 & & \textbf{Ours} & 0.89 & \textbf{29.59} \\
 \cmidrule(lr){2-5}
 & \multirow{4}{*}{0.03} & Baseline & 0.98 & 1211.11 \\
 & & \textbf{Max Score} & 0.94 & \textbf{132.31} \\
 & & Z-Score & 0.93 & 197.77 \\
 & & Ours & 0.97 & 160.66 \\
\midrule

\multirow{12}{*}{FIQA} & \multirow{4}{*}{0.10} & Baseline & 0.89 & 417.77 \\
 & & Max Score & 0.87 & 83.23 \\
 & & Z-Score & 0.87 & 78.02 \\
 & & \textbf{Ours} & 0.86 & \textbf{56.72} \\
 \cmidrule(lr){2-5}
 & \multirow{4}{*}{0.05} & Baseline & 0.94 & 846.0 \\
 & & Max Score & 0.92 & 254.8 \\
 & & Z-Score & 0.92 & 217.79 \\
 & & \textbf{Ours} & 0.92 & \textbf{190.5} \\
 \cmidrule(lr){2-5}
 & \multirow{4}{*}{0.03} & Baseline & 0.96 & 1206.93 \\
 & & Max Score & 0.94 & 380.62 \\
 & & Z-Score & 0.94 & 437.01 \\
 & & \textbf{Ours} & 0.95 & \textbf{347.16} \\

\bottomrule
\end{tabular}
}
\caption{Ablation study comparing different score refinement methods with BGE-large-v1.5 encodings. The table shows empirical coverage and average group size for different datasets and methods. Bold values indicate the best performance for each $\alpha$.}
\label{tab:app_exp}
\end{table}